# COVID-19 detection using ViT transformer-based approach from Computed Tomography Images

Kenan Morani

kenan.morani@gmail.com

Electrical and Electronics Engineering Department, Izmir Democracy University, Izmir, Turkey

0000-0002-4383-5732

**ABSTRACT**

In here, we introduce a novel approach to enhance the accuracy and efficiency of COVID-19 diagnosis using CT images. Leveraging state-of-the-art Transformer models in computer vision, we employed the base ViT Transformer configured for 224x224-sized input images, modifying the output to suit the binary classification task. Notably, input images were resized from the standard CT scan size of 512x512 to match the model's expectations. Our method implements a systematic patient-level prediction strategy, classifying individual CT slices as COVID-19 or non-COVID. To determine the overall diagnosis for each patient, a majority voting approach as well as other thresholding approaches were employed. This method involves evaluating all CT slices for a given patient and assigning the patient the diagnosis that relates to the thresholding for the CT scan. This meticulous patient-level prediction process contributes to the robustness of our solution as it starts from 2D-slices to 3D-patient level. Throughout the evaluation process, our approach resulted in 0.7 macro F1 score on the COV19-CT -DB validation set. To ensure the reliability and effectiveness of our model, we rigorously validate it on the extensive COV-19 CT dataset, which is meticulously annotated for the task. This dataset, with its comprehensive annotations, reinforces the overall robustness of our solution.

1. INTRODUCTION

The severity of the COVID-19 pandemic has spurred a global effort to develop innovative and effective solutions for mitigating the spread of the virus and its early detection. In this context, medical imaging, particularly the use of computed tomography (CT) images, has emerged as a valuable tool for aiding in the diagnosis of COVID-19. The ability to visualize and analyze the impact of the virus on the human respiratory system has played a crucial role in understanding and responding to the pandemic. CT images provide rich visual data that can potentially expedite the identification of affected individuals and inform clinical decision-making, making them an essential asset in the battle against COVID-19 [1].

As the field of medical imaging evolves, so does the demand for state-of-the-art solutions that can effectively analyze and interpret the wealth of visual data available. Vision Transformers (ViTs) have garnered significant attention as a groundbreaking approach in the domain of computer vision and image analysis. Their remarkable success in various vision tasks, including the detection of illnesses and abnormalities, underscores their potential as a transformative technology in the context of COVID-19. By harnessing the power of Vision Transformers, we can enhance the accuracy and efficiency of COVID-19 detection, potentially reducing the spread of the virus and enabling early intervention to improve patient outcomes [2].

In this academic paper, we present a comprehensive study that leverages the advanced capabilities of a base model pretrained ViT Transformer. This state-of-the-art model is tailored to the specific demands of illness detection from CT images and operates with an input image size of 224 [3]. Our research endeavors to push the boundaries of COVID-19 diagnosis and management by harnessing the potential of Vision Transformers in a medical context, marking a significant step forward in the fight against the pandemic.

## 2. METHODOLOGY

### 2.1 The Dataset

The dataset utilized in this study is an extension of the COV19-CT-DB database, encompassing annotated CT scans from a total of 1,650 COVID and 6,100 Non-COVID cases. The meticulous annotation process was carried out by a panel of experts, each possessing over two decades of experience, with four experts in total. Notably, every CT scan comprises a variable number of slices, ranging from 50 to 700.

For the purpose of this research, we retained the original training set and a subset of the initial validation set, resulting in a validation set of 368 COVID cases and 469 Non-COVID cases, as shown in Table 1. This partitioning strategy ensures that the dataset maintains its integrity while facilitating rigorous evaluation of the proposed methodology.

Access to this dataset is made available through the "ECCV 2022: 2nd COV19D Competition". It is important to highlight that the outcomes presented in this paper were derived after the successful submission of an extended version of the database for the IEEE ICASSP 2023: AI-enabled Medical Image Analysis Workshop and Covid-19 Diagnosis Competition (AI-MIA-COV19D). Interested parties may request access to the COV19-CT database directly from the organizers of the workshop [4-5-6-7-8-9-10].

*Table 1 Distribution of cases in training and validation partitions*

| Annotation | Training | Validation |
|---|---|---|
| COVID-19 cases | 687 | 215 |
| Non-COVID cases | 867 | 269 |

### 2.2 The Model Architecture

At the core of the ViT model, the Transformer-based architecture excels in capturing both local and global dependencies within an image, specifically tailored for the `vit_base_patch16_224` variant. The essential concept of a ViT remains unchanged: the input image is divided into fixed-size non-overlapping patches, treated as tokens, and processed through a series of Transformer layers. This patch-based strategy offers scalability and adaptability to handle images of varying sizes. The introduction of positional embeddings ensures the encoding of spatial information, allowing the model to understand relative patch positions. The ViT architecture, comprising multiple transformer blocks, each housing attention mechanisms and feedforward neural networks, collaboratively processes visual data to extract meaningful features.

Moreover, ViTs often integrate a classification head for making predictions based on learned features, enhancing their versatility across various vision tasks. The Transformer, specifically the `vit_base_patch16_224` variant, stands as a cutting-edge architecture, particularly well-suited for medical image analysis, including the detection of illnesses like COVID-19 from CT images.

Classification Head: The final layer, the classification head, produces predictions, configured for binary outcomes in the case of medical image analysis, distinguishing COVID and Non-COVID cases [11-12]. The transformer model's application in medical image analysis represents a state-of-the-art approach, demonstrating potential in enhancing the accuracy and efficiency of illness detection, such as COVID-19, from CT images.

In our study, input images underwent critical preprocessing to align seamlessly with the `vit_base_patch16_224` architecture. Original images from the COV19-CT database were uniformly resized to a resolution of 224x224 pixels. This resizing ensures compatibility with the Transformer's patch-based architecture, facilitating effective visual data processing.

It is crucial to note that, apart from resizing, no further modification or extensive data augmentation was applied to the input images. This approach preserves their essential characteristics and information, emphasizing the utilization of the transformer's inherent power to capture valuable patterns and features within the CT images.

The output of the Transformer was tailored to address the specific objectives of our study, which involved the binary classification of COVID-19 cases and Non-COVID cases. In alignment with this task, the model's final classification head was configured to make precise binary predictions. This binary classification setting is instrumental in facilitating the diagnosis of COVID-19, a pivotal objective of our research, as highlighted in our GitHub repository dedicated to this endeavor.

The model was compiled via pytorch platform with the following tuning:

- Learning Rate (0.001): The learning rate is set to control convergence speed without causing instability.
- Number of Epochs (20): Training for 20 epochs allows the model to converge to an optimal solution.
- Batch Size (32): A batch size of 32 balances training speed and stability.
- Number of Classes (2): The binary classification task involves two classes.
- Loss Function (Cross-Entropy Loss): Cross-entropy loss is suitable for classification tasks.
- Optimizer (Adam): Adam optimizer is efficient for training deep models.

The equipment used is GNU/Linux operating system on 64GiB System memory with Intel(R) Xeon(R) W-2223 CPU @ 3.60GHz processor.

### 2.3 Patient Level Predictions

In order to make patient-level predictions, a systematic approach was implemented, involving the following steps:

1. Data Iteration: The code iterates through the image files within the specified folder path, which contains a collection of CT images. Each folder within the main directory corresponds to an individual patient's CT scan.
2. Prediction at Slice Level: For each CT scan, the code loads and preprocesses the CT images. These images are then passed through the previously trained Transformer model to obtain predictions. The predictions provide a binary classification of each image slice as either COVID-19 or non-COVID, based on the computed class probabilities.
3. Majority Voting: The code accumulates the predictions for all slices within a patient's CT scan. For each patient, the code tallies the count of predicted COVID-19 and non-COVID slices.
4. Patient Label Determination: The patient's label is determined based on the majority of predictions within their CT scan. If the count of predicted COVID-19 slices is higher than that of non-COVID slices, the patient is categorized as COVID-19 positive. Conversely, if the count of predicted non-COVID slices is greater, the patient is labeled as non-COVID. This majority voting method at the patient level enables a robust diagnosis of the patient's COVID-19 status.

By implementing this approach, the code achieves patient-level predictions that take into account the collective information from all slices within a patient's CT scan, providing an effective method for COVID-19 diagnosis based on CT image data.

### 2.4 Performance Evaluation

The proposed model was evaluated via the COV19-CT-DB database using accuracy, macro F1 score and confidence interval.

The accuracy is calculated as in Equation 1:

$$Accuracy = \frac{True\ Positives + True\ Negatives}{True\ Positives + False\ Positives + True\ Negatives + False\ Negatives} \qquad (1)$$

Where positive and negative cases refer to COVID and Non-COVID cases.

The macro F1 score was calculated after averaging precision and recall matrices as in Equation 2:

$$Macro\ F1 = \frac{2 \times average\ precision \times average\ recall}{average\ precision + average\ recall} \qquad (2)$$

Furthermore, to report the confidence intervals of the results obtained, the Binomial proportion confidence intervals for macro F1 score are used. The confidence intervals were used to check the range variance of the reported results. The residuals of the interval can be calculated as in Equation 3 [24].

$$\text{Radius of Interval} = z \times \sqrt{\frac{\text{macro F1} \times (1 - \text{macro F1})}{n}} \qquad (3)$$

where z is the number of standard deviations from the Gaussian distribution and n is the number of samples.

## 3. RESULTS

### 3.1 Results on the validation partition

Table 3 shows the training performance.

*Table 2 Performance results of the training*

| Performance metric | Score |
|---|---|
| Average training accuracy | 75.45% |
| Average recall | 75.84% |
| Average precision | 75.33% |

To calculate the confidence interval for the resulting accuracy, equation 3 was used. In the equation, z is taken as z=1.96 for a significance level of 95%. By that we can calculate the confidence interval for the macro F1 score (approximately 0.78) as in Equation 4:

$$\text{interval} = 1.96 \times \sqrt{\frac{0.75\,(1-0.75)}{106378}} \approx 0.00013 \qquad (4)$$

The number of samples (slices) in the validation set is 106,378. The result from the last equation shows sufficient confidence in the resulting accuracy.

### 3.2 Results on the test partition

Our looked at slices of each CT scan belonging to one patient and decided whether it is a COVID case or Non-COVID based on different voting methods called thresholds. Fig. 1 shows the method's performance results for different thresholding methods. To explain one voting method, we may choose 0.4 thresholds. In here, if the number of COVID slices in one CT scans is greater than 20% of the number of Non-COVID slices in the same CT scan, then the patient is diagnosed with the disease. Else the patient will be healthy. A threshold of 40% gave the highest Validation accuracy and weighted macro F1 score among the tried thresholds

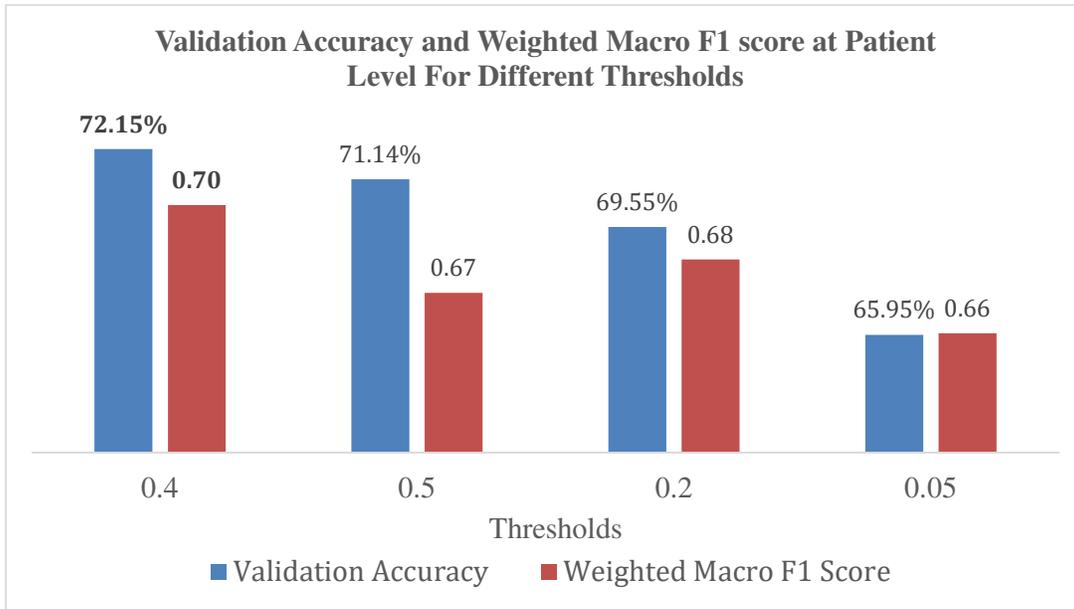

*Fig 1 Validation accuracy and Weighted Macro F1 score at patient level for different thresholds*

The following are the results when using 40% thresholding voting method.

Confusion matrix at patient level is shown in Table 3.

*Table 3 Confusion Matrices at patient level*

|  |  | COVID | NONCOVID |
|---|---|---|---|
|  | **COVID** | 142 | 110 |
| **PREDICTED** | **NONCOVID** | 83 | 358 |

Precision, recall and accuracy at patient level are shown in Table 4.

*Table 4 Precision, Recall and F1 score*

| F1 COVID | 0.60 |
|---|---|
| F1 NONCOVID | 0.79 |
| Macro Average | 0.69 |

Table 5 shows the classe weights in the validation set.

*Table 5 Weighted of COVID and Non-COVID classes in the validation set*

| Weighted F1 COVID | 0.41 |
|---|---|
| Weighted F1 NONCOVID | 0.53 |

Our approach has achieved a macro F1 score that not only exceeds the established baseline but also surpasses the scores achieved by a multitude of other competitors on the validation set of the data. This remarkable performance underscores the effectiveness and superiority of our method in the context of COVID-19 diagnosis using CT images, making it a promising and competitive solution in the field. Table 6 compares our results to some other results presented in the competition [13-14-15].

*Table 6 Comparison with other highest performing alternatives on the COV19-CT Database; validation set in terms of macro F1 score*

| The Method | Validation set | Test set |
|---|---|---|
| Cov3d | 0.94 | 0.878 |
| BERT method | 0.91 | 0.808 |
| Base Line | 0.77 | 0.67 |
| **Our Proposed method** | **0.70** | - |

## 4. CONCLUSION

In this study, we have presented a robust and effective method for COVID-19 diagnosis utilizing CT images. By harnessing the capabilities of transformer models, a cutting-edge technology in computer vision, we have demonstrated the power of deep learning in medical image analysis. Our systematic approach for patient-level predictions, based on individual CT slice classifications and majority voting, extends on our previous work.

The results of our study exhibit advancements in terms of showing the performance using a vision transformer. Future work to improve the performance of the methodology may include, processing the images before inputting them to the transformer. Image processing techniques, such as lung area segmentation may help increase the focus on the interest areas of the CT slices. In addition, different types of transformers can be tested. Replacing the current ViT transformer with a transformer such as Swin transformers or even trying a larger/smaller version of the same transformer may improve the results

## ACKNOWLEDGEMENT

Acknowledgement goes to the medical staff which worked on annotating COV19-CT-DB database and other members who shared the dataset with IDU-CVLab team.

## DECLARATIONS

**Funding statement.** No funding was provided for this study.

**Conflict of interest.** The author declares no conflict of interest.

**Additional information.** The code related to this study can be found on Github at https://github.com/IDU-CVLab/COV19D_4th